\documentclass[12pt]{article}

\usepackage{amsmath}
\usepackage{amssymb}
\usepackage{amsfonts}
\usepackage{latexsym}
\usepackage{bm}
\usepackage{color}

\catcode `\@=11 \@addtoreset{equation}{section}

\catcode `\@=12

\newcommand{\be}{\begin{equation}}
\newcommand{\en}{\end{equation}}
\newcommand{\bea}{\begin{eqnarray}}
\newcommand{\ena}{\end{eqnarray}}
\newcommand{\beano}{\begin{eqnarray*}}
\newcommand{\enano}{\end{eqnarray*}}
\newcommand{\bee}{\begin{enumerate}}
\newcommand{\ene}{\end{enumerate}}

\newcommand{\N}{\mathfrak N}

\newcommand{\mc}{\mathcal}

\newcommand{\Sc}{{\cal S}}
\newcommand{\E}{{\cal E}}
\newcommand{\F}{{\cal F}}

\newcommand{\1}{1 \!\! 1}

\newcommand{\kt}{\rangle}
\newcommand{\br}{\langle}

\newcommand{\Hil}{\mc H}

\catcode `\@=11 \@addtoreset{equation}{section}
\catcode `\@=12

\begin{document}

\thispagestyle{empty}

\vspace*{2cm}

\begin{center}
{{\Large \bf Some results on the dynamics and  transition probabilities for non self-adjoint hamiltonians}}\\[10mm]


{\large F. Bagarello} \footnote[1]{ Dipartimento di Energia, Ingegneria dell'Informazione e Modelli Matematici,
Facolt\`a di Ingegneria, Universit\`a di Palermo, I-90128  Palermo, and INFN, Universit\`a di di Torino, ITALY\\
e-mail: fabio.bagarello@unipa.it\,\,\,\, Home page: www.unipa.it/fabio.bagarello}


\end{center}

\vspace*{2cm}

\begin{abstract}
\noindent We discuss systematically several possible inequivalent ways to describe the dynamics and the transition probabilities of a quantum system when its hamiltonian  is not self-adjoint.  In order to simplify the treatment,  we mainly restrict our analysis to finite dimensional Hilbert spaces.  In particular, we propose some experiments which could discriminate between the various possibilities considered in the paper. An example taken from the literature is discussed in detail.

\end{abstract}


\vfill


\newpage

\section{Introduction}

In ordinary quantum mechanics one of the fundamental axiom of the whole theory is that the hamiltonian $H$ of the physical system must be self-adjoint: $H=H^\dagger$. This condition, shared also by all the {\em observables} of the system, is important since it ensures that the eigenvalues of these observables, and of the hamiltonian in particular, are real quantities. However, this is not a necessary condition, and in fact several physically motivated examples exist in the literature concerning non self-adjoint operators whose spectra consist of only real eigenvalues.

However, $H=H^\dagger$ has an extra bonus, since the time evolution deduced out of $H$ is unitary and, being so, preserves the total probability: if $\Psi(t)$ is a solution of the Schr\"odinger equation $i\dot\Psi(t)=H\Psi(t)$, then $\|\Psi(t)\|^2$ does not depend on time. This is clear since $\Psi(t)=e^{-iHt}\Psi(0)$, and since $U_t=e^{-iHt}$ is unitary. Of course, this is false if $H\neq H^\dagger$, and in fact, in this case,  $\|\Psi(t)\|^2$ does indeed depend on time, in general. Sometimes this is exactly what one looks for: in many simple systems in quantum optics, for instance, non self-adjoint hamiltonians are used to describe some decay, so that there is no reason for the probability to be preserved in time. Other times, one would prefer to avoid any damping, so that the aim is to find some way to {\em recover unitarity} even when $H\neq H^\dagger$. This is particularly interesting for people in the PT-community, who quite often work with hamiltonian operators which are not self-adjoint, but simply pseudo-symmetric or PT-symmetric, \cite{ben,ali}, and in fact several attempts have been proposed along the years by different authors to discuss this and other aspects of time evolution for systems driven by non self-adjoint hamiltonians. Here we refer to \cite{brody}-\cite{mosta3}, and references therein.  However, in our opinion, much more can be said, and using a rather general approach. This is exactly what we will do here, in the next section, considering the cases in which the eigenvalues of $H$ are all real and commenting on the situation in which some eigenvalues are complex.

In all this paper we will work with finite-dimensional Hilbert spaces. This has two nice consequences: the first one is that all the operators involved are bounded (hence, everywhere defined) and the inverse, when it exists, is bounded as well. In fact, we are dealing with matrices. The second consequence is that we can easily, quite often, discuss examples in terms of pseudo-fermions (PFs), \cite{bagpf,bagpf2}, as we have already recently shown in \cite{baggar}. We should stress that, contrarily to what often stated in the literature, going from a finite to an infinite dimensional Hilbert space is an absolutely non trivial task. Therefore, most of our claims, though giving indications also in this latter case, are rigorously true only in the present, finite-dimensional, settings. We will comment more on this aspect all along the paper.

This article is organized as follows:

In the next section we discuss the general functional structure associated to a non self-adjoint hamiltonian, and its dynamics. We also comment briefly on the case of non purely real eigenvalues and on finite temperature equilibrium states. In Section 3 we propose different definitions of transition probability functions, and we discuss a possible strategy to discriminate between them. This is, in fact, the core of our paper since it could be used, in principle, to deduce which are the {\em correct} Hilbert space, scalar product, norm and adjoint, or, more explicitly, which definitions reproduce the experimental data. This proposal is made more precise in Section 4, with the aid of an explicit example, originally introduced in \cite{santos} and discussed here adopting a simple and general pseudo-fermionic representation. Section 5 contains our conclusions. To keep the paper self-consistent, we list some definitions and results on PFs in the Appendix.

\section{A general settings for $H\neq H^\dagger$}

As we have already said, in this paper we will focus on the easiest situation, i.e. on finite dimensional Hilbert spaces. In this way our operators are finite matrices. The main ingredient is an operator (i.e. a matrix) $H$, acting on the vector space ${\Bbb C}^{N+1}$, with $H\neq H^\dagger$ and with exactly $N+1$ distinct eigenvalues $E_n$, $n=0,1,2,\ldots,N$. Here, the adjoint $H^\dagger$ of $H$ is the usual one, i.e. the complex conjugate of the transpose of the matrix $H$. Because of what follows, and in order to fix the ideas, it is useful to remind here that the adjoint of an operator $X$, $X^\dagger$, is defined in terms of the {\em natural} scalar product $\left<.,.\right>$ of the Hilbert space $\Hil=\left({\Bbb C}^{N+1},\left<.,.\right>\right)$:
$\left<Xf,g\right>=\left<f,X^\dagger g\right>$, for all $f,g\in {\Bbb C}^{N+1}$, where $\left<f,g\right>=\sum_{k=0}^N\overline{f_k}\,g_k$, with obvious notation. We will consider separately the case in which all the eigenvalues $E_n$ are real and the situation in which some are complex. In both cases we will assume that each $E_n$ has multiplicity one.

Before starting, it is necessary to clarify some notation adopted in this paper: we will use ${\Bbb C}^{N+1}$ any time we want to stress the nature of vector space of our vectors. When it is important to stress the topological (i.e. the scalar products and the norms) aspects of this set, we will use $\Hil$ instead of ${\Bbb C}^{N+1}$ (and, later, $\Hil_\varphi$ or $\Hil_\Psi$). Before starting with our analysis, it is surely worth stressing that, with a different language, some of the results discussed in Section 2 can be found in the literature, see \cite{ali,mosta1,mosta2,mosta3,milos2} for instance.  We have decided to include these statements here for several reasons: first, they are useful to fix our notation. Secondly, some of the proofs discussed here are different, or cannot be found, in the existing literature. Last but not least, we want to keep an eye to possible extensions of our results to the situation in which infinite dimensional Hilbert spaces are needed, where unbounded operators most probably appear, with all their delicate mathematical aspects.

\subsection{All the eigenvalues are real}

We assume here that $H$ has $N+1$ distinct real eigenvalues, corresponding to $N+1$ eigenvectors $\varphi_k$, $k=0,1,2,\ldots,N$:
\be
H\varphi_k=E_k\varphi_k.
\label{21}\en
The set $\F_\varphi=\{\varphi_k,\,k=0,1,2,\ldots,N\}$ is a basis for ${\Bbb C}^{N+1}$, since the eigenvalues are all different. Then an unique biorthogonal basis of $\Hil$, $\F_\Psi=\{\Psi_k,\,k=0,1,2,\ldots,N\}$, surely exists, \cite{you,chri}: $\left<\varphi_k,\Psi_l\right>=\delta_{k,l}$, for all $k, l$. Moreover, for all $f\in\Hil$, we can write
$f=\sum_{k=0}^N\left<\varphi_k,f\right>\Psi_k = \sum_{k=0}^N\left<\Psi_k,f\right>\varphi_k$. Incidentally, this means that both $\F_\varphi$ and $\F_\Psi$ are complete (or total): if $f\in\Hil$ is such that $\left<\varphi_k,f\right>=0$, or $\left<\Psi_k,f\right>=0$, for all $k$, then $f=0$.

What is interesting for us is that the set $\F_\Psi$ is automatically a set of eigenstates of $H^\dagger$ with eigenvalues $E_k$:
\be
H^\dagger\Psi_k=E_k\Psi_k,
\label{22}\en
$k=0,1,2,\ldots,N$. This follows from the completeness of $\F_\varphi$ and from the following equality, where $k$ is arbitrary but fixed:
$$
\left<\left(H^\dagger\Psi_k-E_k\Psi_k\right),\varphi_l\right>=\left<H^\dagger\Psi_k ,\varphi_l\right>
-\left<E_k\Psi_k ,\varphi_l\right>= \left<\Psi_k ,H\varphi_l\right>
-E_k \left<\Psi_k ,\varphi_l\right>=\left(E_l-E_k\right)\delta_{k,l}=0,
$$
for all $l=0,1,2,\ldots,N$. Then (\ref{22}) follows.

Using the bra-ket notation we can write $\sum_{k=0}^N|\varphi_k\left>\right<\Psi_k|=\sum_{k=0}^N|\Psi_k\left>\right<\varphi_k|=\1$, where, for all $f,g,h\in\Hil$, we define $(|f\left>\right<g|)h:=\left<g,h\right>f$. We introduce the operators $S_\varphi=\sum_{k=0}^N|\varphi_k\left>\right<\varphi_k|$ and $S_\Psi=\sum_{k=0}^N|\Psi_k\left>\right<\Psi_k|$, as in \cite{bagbook}. These are bounded positive, self-adjoint, invertible operators, one the inverse of the other: $S_\Psi=S_\varphi^{-1}$. We want to stress again that, in our present settings, there is absolutely no problem with the domains of these (and other) operators, while in \cite{bagbook} we have discussed what happens for infinite-dimensional Hilbert spaces. This passage is absolutely non trivial, but will not be considered here\footnote{Operators of this kind have been introduced by several authors, with different names, in recent years. Some references are \cite{ali,milos2,petr}, where other references can be found. However, only recently the relevance of unbounded operators in this context has been recognized, \cite{petr,sergei,bagzno2,mosta4}. As it is well known, in this case several problems, mainly related to domain problems, must be taken into account.}. Using standard techniques in functional analysis, or direct matrix computations, we can introduce the positive square roots of $S_\Psi$ and $S_\varphi$, and again we have $S_\Psi^{1/2}=S_\varphi^{-1/2}$. Useful (and well known) properties are the following:
\be
S_\varphi\Psi_n=\varphi_n,\quad S_\Psi\varphi_n=\Psi_n,\quad \mbox{ as well as }\quad S_\Psi H=H^\dagger S_\Psi,\quad S_\varphi H^\dagger= HS_\varphi.
\label{23}\en
 If we now define $H_0:=S_\Psi^{1/2}HS_\varphi^{1/2}$ and $e_k=S_\Psi^{1/2}\varphi_k= S_\varphi^{1/2}\Psi_k$, $k=0,1,2,\ldots,N$, we see that $H_0=H_0^\dagger=S_\varphi^{1/2}H^\dagger S_\Psi^{1/2}$,  and that $\E=\{e_k,\,k=0,1,2,\ldots,N\}$ is an orthonormal (o.n.) basis of $\Hil$ of eigenstates of $H_0$: $H_0e_k=E_ke_k$.

Similarly to what is done in many places in the literature, these operators can be used now to define new scalar products in ${\Bbb C}^{N+1}$:
\be
\left<f ,g\right>_\varphi:=\left<S_\varphi^{1/2}f ,S_\varphi^{1/2}g\right>=\left<S_\varphi f ,g\right>, \qquad
\left<f ,g\right>_\Psi:=\left<S_\Psi^{1/2}f ,S_\Psi^{1/2}g\right>=\left<S_\Psi f ,g\right>,
\label{24}\en
for all $f,g\in{\Bbb C}^{N+1}$. Due to the properties of $S_\Psi$ and $S_\varphi$, these are scalar products, everywhere defined on ${\Bbb C}^{N+1}$. Of course, the related norms $\|.\|$, $\|.\|_\varphi$ and $\|.\|_\Psi$ are all equivalent\footnote{This means that, if a sequence of vectors $f_n\in {\Bbb C}^{N+1}$ converges in $\|.\|$, it also converges in $\|.\|_\varphi$ and in $\|.\|_\Psi$.}. For instance we can check that
$$
\frac{1}{\|S_\varphi^{-1/2}\|}\,\|f\|\leq\|f\|_\varphi\leq \|S_\varphi^{1/2}\|\,\|f\|,
$$
for all $f\in{\Bbb C}^{N+1}$. Then, from a topological point of view, $\Hil$, $\Hil_\varphi:=\left({\Bbb C}^{N+1},\left<.,.\right>_\varphi\right)$ and $\Hil_\Psi:=\left({\Bbb C}^{N+1},\left<.,.\right>_\Psi\right)$ are all equivalent. However, they are different under other aspects, as we will show in a moment. In particular, the first difference is in the definition of the adjoint of the operators, which is $\dagger$ in $\Hil$, but which becomes $\flat$ in $\Hil_\varphi$ and $\sharp$ in $\Hil_\Psi$: $\left<Xf,g\right>_\varphi=\left<f,X^\flat g\right>_\varphi$ and $\left<Xf,g\right>_\Psi=\left<f,X^\sharp g\right>_\Psi$, for all $f,g\in {\Bbb C}^{N+1}$. It is easy to see that $\sharp$ and $\flat$ are really adjoints\footnote{For instance, $(X^\sharp)^\sharp=X$ and $(XY)^\sharp=Y^\sharp X^\sharp$, for all operators $X$ and $Y$ on ${\Bbb C}^{N+1}$.}, and to deduce the following relations:
\be
X^\flat=S_\Psi X^\dagger S_\varphi,\qquad X^\sharp=S_\varphi X^\dagger S_\Psi\qquad \mbox{and}\quad X^\flat=S_\Psi^2 X^\sharp S_\varphi^2,
\label{2add1}\en
for each operator $X$ on ${\Bbb C}^{N+1}$. It is now an easy computation to check that $H=H^\sharp$, and that $H^\dagger=(H^\dagger)^\flat$. Indeed we have $\left<Hf,g\right>_\Psi=\left<f,Hg\right>_\Psi$ and $\left<H^\dagger f,g\right>_\varphi=\left<f,H^\dagger g\right>_\varphi$, for all $f,g\in {\Bbb C}^{N+1}$.

\vspace{2mm}

{\bf Remark:--} the equalities in (\ref{2add1}) cannot be extended easily if $\dim(\Hil)=\infty$. The reason is the following: if, for instance, $X^\dagger$ and $S_\varphi$ are unbounded, taken $f\in D(S_\varphi)$, the domain of $S_\varphi$, there is no reason a priori for $S_\varphi f$ to belong to $D(X^\dagger)$, so that $X^\dagger S_\varphi f$ needs not to be defined.  Therefore, when $\dim(\Hil)=\infty$, the three Hilbert spaces are different not only topologically, but also as sets. Of course, this cannot happen if $\dim(\Hil)<\infty$, since all the operators can be defined in all of $\Hil$.

\subsubsection{The dynamics}

The aspect we are interested in here is the dynamics associated to $H$ and to $H^\dagger$, since there is no reason a-priori  for the Schr\"odinger equation to hold also in this case. This problem has been discussed by many authors over the past decade, see \cite{nesterov,milos1,mosta1,mosta3} for instance, with a particular attention, sometimes to the unitarity of the time evolution. However, our point of view is slightly different  since we are more interested in: (1) motivating the {\em extended Schr\"odinger equation} usually taken for granted in the literature, and (2) trying to understand if it is possible to deduce which is the {\em best choice} (with {\em best} to be somehow defined!) for the time evolution of a system, in this case.

Our starting assumption is that, since $H_0$ is self-adjoint, it {\em produces} a standard Schr\"odinger equation for the wave function $\Phi(t)$: $i\dot\Phi(t)=H_0\Phi(t)$, whose solution is $\Phi(t)=e^{-iH_0t}\Phi(0)$. This function satisfies the following equality: $p(t):=\|\Phi(t)\|^2=\left<\Phi(t),\Phi(t)\right>=p(0)$. Then, probability is preserved. What is interesting is that, using our results, and in particular the similarity equations $H_0:=S_\Psi^{1/2}HS_\varphi^{1/2}=S_\varphi^{1/2}H^\dagger S_\Psi^{1/2}$,
 two related Schr\"odinger-like equations naturally arise, with simple computations, from the one for $\Phi(t)$:
\be
i\dot\Phi_\Psi(t)=H\Phi_\Psi(t), \qquad\mbox{and}\qquad i\dot\Phi_\varphi(t)=H^\dagger\Phi_\varphi(t),
\label{25}\en
where $\Phi_\Psi(t)=S_\Psi^{-1/2}\Phi(t)$ and $\Phi_\varphi(t)=S_\varphi^{-1/2}\Phi(t)$. Of course, both these equations can be easily solved:
\be
\Phi_\Psi(t)=e^{-iHt}\Phi_\Psi(0)=e^{-iHt}S_\Psi^{-1/2}\Phi(0), \qquad \Phi_\varphi(t)=e^{-iH^\dagger t}\Phi_\varphi(0)=e^{-iH^\dagger t}S_\varphi^{-1/2}\Phi(0).
\label{26}\en
Now, the equations in (\ref{25}) suggest that $H$ and $H^\dagger$ produce, by themselves, two  Schr\"odinger-like equations which could be, therefore, taken as the \underline{starting points} to describe the dynamics of the given system. This is, in fact, what it is done, quite often, in the literature, \cite{nesterov,milos1,milos2} etc.: one uses a non self-adjoint hamiltonian $H$, and simply writes the equation of the dynamics for the wave function $\xi(t)$ as in (\ref{25}): $i\dot\xi(t)=H\xi(t)$. The two equations in (\ref{25}), which are deduced directly from $i\dot\Phi(t)=H_0\Phi(t)$, suggest that this is in fact reasonable. Of course, the {\em old} probability is not preserved for $\Phi_\Psi(t)$ and $\Phi_\varphi(t)$, at least if we stick with the norm $\|.\|$. But there is no a-priori reason to adopt this particular choice. One could use $\|.\|_\varphi$ or $\|.\|_\Psi$, and indeed we have
$$
p_\varphi(t):=\|\Phi_\varphi(t)\|_\varphi^2=\left<\Phi_\varphi(t),\Phi_\varphi(t)\right>_\varphi=p_\varphi(0)
$$
and
$$
p_\Psi(t):=\|\Phi_\Psi(t)\|_\Psi^2=\left<\Phi_\Psi(t),\Phi_\Psi(t)\right>_\Psi=p_\Psi(0),
$$
for all $t\in\Bbb R$. Then we see that, if for some reason we are interested in having the probability associated to $\Phi_\varphi(t)$ preserved in time, the natural framework to use is not the one provided by $\Hil$ but that given by $\Hil_\varphi$. We will go back to this point in Section 3.

\vspace{2mm}

This freedom of choice is reflected by the dynamics of the operators in the Heisenberg representation: in standard quantum mechanics, i.e. working with $H_0=H_0^\dagger$ in $\Hil$, the recipe to deduce the time evolution of the observable $\hat X$ is simple: its time evolution, $\hat X(t)$, should have on the vector $\xi$ describing the physical system at $t=0$ the same mean value of $\hat X$ on the time-evoluted vector $\xi(t)=e^{-iH_0t}\xi$:
$$
\left<\xi,\hat X(t)\xi\right>=\left<\xi(t),\hat X\xi(t)\right>=\left<e^{-iH_0t}\xi,\hat Xe^{-iH_0t}\xi\right>=\left<\xi,e^{iH_0t}\hat Xe^{-iH_0t}\xi\right>.
$$
Then $\hat X(t)=e^{iH_0t}\hat Xe^{-iH_0t}$. If we repeat the same steps, replacing $H_0$ with $H$, we deduce that
\be
\hat X_\Hil(t)=e^{iH^\dagger t}\hat Xe^{-iHt}.
\label{add1}\en
But this is indeed not the only possibility, and it is not even the most convenient, since during this time evolution we lose three crucial characteristics of the {\em standard} time evolution: (1) the operators $e^{iH^\dagger t}$ and $e^{-iH t}$ are no longer unitary. Sometimes, however, this is exactly what one looks for. As already stressed, this is what is done, for instance, in quantum optics, to describe some damping, \cite{benaryeh,tripf}; (2) An annoying consequence of (\ref{add1}) is that it is not so easy to find integrals of motion for the system, since $[H,\hat X]=0$ does not imply that $\hat X(t)=\hat X(0)$. Then, it is no longer true that the observables which commute with $H$ do not evolve in time; (3) A more serious difficulty is that the time evolution is no longer an automorphism of the set of observables, since in general $(\hat X\hat Y)_\Hil(t)\neq \hat X_\Hil(t)\hat Y_\Hil(t)$, and this complicates in an enormous way all the computations. For instance, in general, it is absolutely non trivial to deduce the analogous of the Heisenberg equation of motion  for $\hat X(t)$.

However, there is a possible way out, and it has again to do with a suitable choice of the scalar product. In fact, if we work with $\Hil_\Psi$ rather than with $\Hil$, which is a natural choice since $H$ is self-adjoint in $\Hil_\Psi$, we can define a different map: $t\rightarrow \hat X_{\Hil_\Psi}(t)$ working as before:
$$
\left<\xi_\Psi(t),\hat X\xi_\Psi(t)\right>_\Psi=\left<\xi_\Psi,\hat X_{\Hil_\Psi}(t)\xi_\Psi\right>_\Psi,
$$
and the result is the following:
\be
\hat X_{\Hil_\Psi}(t)=e^{iHt}\hat X e^{-iHt},
\label{27}\en
and problems (1), (2) and (3) are solved, paying the only price to work in $\Hil_\Psi$, i.e. to replace $\left<.,.\right>$ with $\left<.,.\right>_\Psi$. Apparently, this is not a big price, indeed. This is not yet the end of the story, since one may wonder why $H$ should be {\em better} than $H^\dagger$. In fact, there is essentially no difference, a priori. In fact, if for some reason we need to use $H^\dagger$ rather than $H$, the choice of working in $\Hil$ would create the same kind of problems (1), (2) and (3) as before. The possible way out is now clearly to use $\Hil_\varphi$ as the natural Hilbert space to work with. In this case, the time evolution of $\hat X$ looks like
\be
\hat X_{\Hil_\varphi}(t)=e^{iH^\dagger t}\hat X e^{-iH^\dagger t},
\label{28}\en
and again those problems are solved.

Summarizing, we can say that, under our assumptions on the $E_n$, working with a non self-adjoint hamiltonian $H$ gives a lot of freedom: we have three Hilbert spaces, with their scalar products, their norms and their involutions, and we further have several possible definitions of the time evolution of any observable, each one with pros et contra. Of course, we are left with a very natural, and deep, question: how should we choose the {\em right} Hilbert space? Or, even better, who or what decides what is {\em right}? We will suggest a possible way to answer these questions in Sections 3 and 4. This is a problem which, in our knowledge, has not been considered in details in the literature so far.

\vspace{2mm}

{\bf Remarks:--} (1) Quite interestingly, and as it is implicit in the work of several  authors, see \cite{ali,brody,milos1} among others, in this context it is the hamiltonian $H$ itself which somehow fixes its preferred Hilbert space! This is because both $\left<.,.\right>_\varphi$ and $\left<.,.\right>_\Psi$ are defined via $S_\varphi$ and $S_\Psi$, which are deduced, in turns, by the eigenvectors of $H$ and $H^\dagger$. This is similar to what happens in algebraic quantum dynamics, see \cite{bagrev} and references therein, where the hamiltonian (self-adjoint, in that context) is used to define a suitable topology on the algebra of the operators needed in the description of the physical system.

\vspace{1mm}

(2) Of course, the possibility of having different definitions for the time evolution is reflected in the definition of the equilibrium states for non-zero temperature. More explicitly, if we use $H_0$ to define the dynamics as $X(t)=e^{iH_0t}Xe^{-iH_0T}$, the natural choice of equilibrium state is the Gibbs state
$$
\omega_0(X):=\frac{1}{Z_0}\,tr(e^{-\beta H_0}X),
$$
where $Z_0=tr(e^{-\beta H_0})$  $\beta=\frac{1}{T}$, $T$ being the temperature of the system, and $tr(A)$ is the trace of the operator $A$. Here we are fixing to one the Boltzmann's constant $K$. Of course, this is not the most reasonable choice if one imagine that the time evolution of the observable $X$ is given by $X_{\Hil_\Psi}(t)=e^{iH t}\hat X e^{-iH t}$ or by $X_{\Hil_\varphi}(t)=e^{iH^\dagger t}\hat X e^{-iH^\dagger t}$. To these choices, in fact, it is more natural to associate the following states
$$
\omega_\Psi(X):=\frac{1}{Z_\Psi}\,tr(e^{-\beta H}X),\qquad \omega_\varphi(X):=\frac{1}{Z_\varphi}\,tr(e^{-\beta H^\dagger}X),
$$
where $Z_\Psi=tr(e^{-\beta H})$ and $Z_\varphi=tr(e^{-\beta H^\dagger})$. To avoid mathematical difficulties, we are thinking as before that the Hilbert space of our system is finite dimensional, so that all the quantities introduced here are well defined. It is not hard to check that all these states satisfy a KMS-like equilibrium condition like, for instance,
$$
\omega_\sharp(A_\sharp(t)B)=\omega_\sharp(BA_\sharp(t+i\beta)),
$$
for all observables $A$ and $B$ and for all possible choices of $\sharp$, i.e. whenever we use the same hamiltonian to define both the time evolution of the system and the state. Hence, loosing self-adjointness does not imply many changes from this point of view. However, it could be interesting to check what happens adopting, for instance, $H_0$ to define the time evolution and, a different but somehow related operator $H$, to define the state.

\subsection{Not all the eigenvalues are real}\label{sectnatear}

Let us now briefly consider what happens when we abandon the assumption that $E_k$ is real for all $k$. This might be relevant for including, in our scheme, some {\em effective} hamiltonians used in different, usually non conservative, contexts, see for instance\cite{benaryeh,tripf} for an application to quantum optics. In this case, equation (\ref{22}) must be replaced by
\be
H^\dagger\Psi_k=\overline{E_k}\,\Psi_k,
\label{29}\en
$k=0,1,2,\ldots,N$, which might seem to be a minor difference. This is not so. In fact, in this new situation, $H$ and $H^\dagger$ are no longer isospectral, and for this reason no invertible intertwining operator exists between $H$ and $H^\dagger$. This makes the framework of the system {\em a little  poorer} than before. In fact, the existence of intertwining operators proved to be quite important in many physical systems, as discussed in many papers on this subject, \cite{intop}.  Even more: suppose $H\neq H^\dagger$ and suppose that $\Im(E_n)\neq0$ for at least one $n$. Then it is easy to show that, contrarily to what happens for real eigenvalues, there exists no scalar product $\ll.,.\gg$ on ${\Bbb C}^{N+1}$ such that $\ll Hf,g\gg=\ll f,Hg\gg$. The proof is trivial, but in our opinion is worth giving: since  $\ll H\varphi_n,\varphi_n\gg=\overline{E_n}\ll \varphi_n,\varphi_n\gg$ and $\ll \varphi_n,H\varphi_n\gg=E_n\ll \varphi_n,\varphi_n\gg$, it is clear that, if $\overline{E_n}\neq E_n$,  $\ll H\varphi_n,\varphi_n\gg\neq \ll \varphi_n,H\varphi_n\gg$. Hence $H$ is not self-adjoint with respect to this different scalar product. The conclusion follows from the arbitrariness of $\ll.,.\gg$.

Then what we did before is no longer true in this new settings: more in details, $\F_\varphi$ and $\F_\Psi$ are still biorthogonal bases (or, most probably, biorthogonal sets if $\dim(\Hil)=\infty$), and $S_\varphi$ and $S_\Psi$ are again one the inverse of the other, positive and self-adjoint. Their square roots exist, and $e_n=S_\Psi^{1/2}\varphi_n$ is still an eigenvector of $H_0=S_\Psi^{1/2}HS_\varphi^{1/2}$ with eigenvalue $E_n$. Also, $\E=\{e_n\}$ is a basis for $\Hil$. However, $H_0\neq H_0^\dagger$, $S_\Psi H\neq H^\dagger S_\Psi$, and $H_0^\dagger e_n=\overline{E_n}e_n$. In particular, this shows that $S_\Psi$ is no longer an intertwining operators between $H$ and $H^\dagger$.

A consequence of these facts is that, assuming as before $i\dot\Phi_\Psi(t)=H\Phi_\Psi(t)$ and $i\dot\Phi_\varphi(t)=H^\dagger\Phi_\varphi(t)$, their solutions do not preserve probabilities. So, in this case, there is no leading rule to follow, apparently.

\vspace{2mm}

{\bf Remark:--} It is clear that when $\Hil$ is finite dimensional, the presence of complex-valued eigenvalues makes not a real big (technical) difference: in fact, we are simply dealing with $(N+1)\times(N+1)$ matrices! However, apart from what we have already discussed, we can expect more problems when $\Hil$ is infinite dimensional. The analysis of this situation is work in progress.

\section{Transition probabilities and consequences}\label{sectTPandC}

In this section we propose three different definitions of transition probabilities, and we deduce the different results which are obtained out of these definitions, suggesting some experiments which make it possible to discriminate among them. In this way we should be able to decide which is the correct framework to adopt when non self-adjoint hamiltonians, with real eigenvalues, are those who naturally describe a physical system $\Sc$.

We first briefly recall what happens in the {\em standard} situation, i.e. when the dynamics of $\Sc$ is driven by a self-adjoint hamiltonian $H_0$, with an o.n. basis $\F_e=\{e_k, \,k=0,1,\ldots,N\}$ of eigenvectors: $H_0\,e_k=E_k\,e_k$, $\forall k$. In this case, $E_k\in\Bbb R$ automatically, for all $k$. If $\Phi_0=\sum_{k=0}^Nc_ke_k$ is the state of the system $\Sc$ at $t=0$, $c_k=\left<e_k,\Phi_0\right>$, then its time evolution is clearly given by $\Phi(t)=e^{-iH_0t}\Phi_0=\sum_{k=0}^Nc_ke^{-iE_kt}e_k$. Here it is not a major request assuming that $\Phi(t)$ is normalized for any $t$, since $e^{-iH_0t}$ is an unitary operator. In order to compute the transition probability to a final state, described by the normalized vector $\Phi_f$, we just need to compute
\be
P_{\Phi_0\rightarrow\Phi_f}(t):=\left|\left<\Phi_f,\Phi(t)\right>\right|^2,
\label{31}\en
which, because of the Schwarz inequality, is clearly always between zero and one: $P_{\Phi_0\rightarrow\Phi_f}(t)\in[0,1]$, for all $t$. Of course, this is strongly related to the fact that both $\Phi_f$ and $\Phi(t)$ are normalized in $\Hil$, and this is possible for all $t\in\Bbb R$ since $H_0$ is self-adjoint. Just as an introductory example, let us now see what happens if $\Phi_0=\varphi_0$ and $\Phi_f=\Psi_0$, where $\varphi_0$ and $\Psi_0$ are two vectors of two generic biorthogonal sets $\F_\varphi=\{\varphi_n\}$ and $\F_\Psi=\{\Psi_n\}$, none of which made of eigenstates of $H_0$. Of course, we are interested in this particular situation in view of our next extension to non self-adjoint hamiltonians. In this case, since normalization is decided by the condition $\left<\varphi_0,\Psi_0\right>=1$, there is no guarantee that $\|\varphi_0\|=\|\Psi_0\|=1$. For this reason, formula (\ref{31}) produces
\be
P_{\Phi_0\rightarrow\Phi_f}(t)=\frac{1}{\|\varphi_0\|^2\|\Psi_0\|^2}\left|\left<\Psi_0,e^{-iH_0t}\varphi_0\right>\right|^2=\frac{1}{\|\varphi_0\|^2\|\Psi_0\|^2}\left|\sum_{k=0}^Nd_k\overline{p_k}e^{-iE_kt}\right|^2,
\label{32}\en
where $d_k=\left<e_k,\varphi_0\right>$ and $p_k=\left<e_k,\Psi_0\right>$. This formula shows that, in particular, $P_{\Phi_0\rightarrow\Phi_f}(t)$ does not depend on time if all the pairs $(d_k,p_k)$, except one, are zero. Also, if the eigenvalues $E_k$ are commensurable, $P_{\Phi_0\rightarrow\Phi_f}(t)$ is a periodic function. Not surprisingly, if $\varphi_0=\Psi_0$, we get $P_{\Phi_0\rightarrow\Phi_f}(0)=1$.

\vspace{3mm}

Let us now consider the non self-adjoint settings described in Section 2. Hence we have $H\neq H^\dagger$, and two sets $\F_\varphi=\{\varphi_n\}$ and  $\F_\Psi=\{\Psi_n\}$ of biorthogonal eigenvectors of $H$ and $H^\dagger$: $H\varphi_n=E_n\varphi_n$, $H^\dagger\Psi_n=E_n\Psi_n$, $\left<\varphi_k,\Psi_n\right>=\delta_{k,n}$. Notice that here we are restricting to real eigenvalues, $E_n\in\Bbb R$, as we will do in all this section.

As we have discussed before, it is natural to assume that the wave function of the system $\Sc$ described by $H$ satisfies the equation $i\dot\Phi(t)=H\Phi(t)$, with initial condition $\Phi(0)=\Phi_0$. Then, since the $\varphi_n$'s are eigenstates of $H$, the natural choice to compute $\Phi(t)$ is to expand this unknown function in terms of $\F_\varphi$. Hence we get
\be
\Phi(t)=e^{-iHt}\Phi_0=\sum_{k=0}^Nc_ke^{-iE_kt}\varphi_k,
\label{33}\en
where $c_k=\left<\Psi_k,\Phi_0\right>$. It is clear that $e^{-iHt}$ is no longer a unitary operator, so that there is no reason for $\|\Phi(t)\|$ to be always equal to one, even when $\|\Phi_0\|=1$. For this reason, at least if we use the norm $\|.\|$, a time dependent normalization must necessarily appear into the game. As a matter of fact, more than one such normalization will now be considered, each one related to a possible different definition of the transition probability in the present context. We should also stress that the necessity of introducing some normalization in similar contexts was already clear to many authors, already several years ago, \cite{sgh}, and recently reconsidered by other authors, \cite{santos}, as well as by many others.

In practice, going from self-adjoint to non self-adjoint hamiltonians opens the possibility of having several possible (apparently) inequivalent definitions of transition probabilities, all of which appear to be absolutely reasonable. The ones we introduce here are the following:
\be
P_{\Phi_0\rightarrow\Phi_f}(t):=\left|\frac{\left<\Phi_f,\Phi(t)\right>}{\|\Phi_f\|\|\Phi(t)\|}\right|^2, \, P_{\Phi_0\rightarrow\Phi_f}^\Psi(t):=\left|\frac{\left<\Phi_f,\Phi(t)\right>_\Psi}{\|\Phi_f\|_\Psi\|\Phi(t)\|_\Psi}\right|^2, \, P_{\Phi_0\rightarrow\Phi_f}^\varphi(t):=\left|\frac{\left<\Phi_f,\Phi(t)\right>_\varphi}{\|\Phi_f\|_\varphi\|\Phi(t)\|_\varphi}\right|^2,
\label{34}\en
which are the most plausible definitions, but not the only ones, as we will discuss later.
They look different since they involve different scalar products and different norms. We will see that there is more than this: they really produce different results, so that \underline{they are not physically} \underline{equivalent at all} and it should be possible, in principle, to discriminate among them, in order to understand which is the most appropriate expression of the transition probability, and why. Incidentally we observe that, with these definitions, $P_{\Phi_0\rightarrow\Phi_f}^\sharp(t)\in[0,1]$ for all $t\geq0$. Here $P_{\Phi_0\rightarrow\Phi_f}^\sharp$ stands for $P_{\Phi_0\rightarrow\Phi_f}$, $P_{\Phi_0\rightarrow\Phi_f}^\Psi$ or $P_{\Phi_0\rightarrow\Phi_f}^\varphi$.

\vspace{2mm}

{\bf Remark:--} At a first sight, $P_{\Phi_0\rightarrow\Phi_f}^\Psi$ may appear as the more natural choice, since $H$ is self-adjoint with respect to $\left<.,.\right>_\Psi$, and therefore $e^{-iHt}$ is unitary with respect to this scalar product. However, here we are not considering unitarity of the time evolution as our main requirement. We are much more interested in a comparison between our theoretical results and some experimental data, and a good agreement is not necessarily ensured by the unitarity of $e^{-iHt}$. Notice also that, if we replace $H$ with $H^\dagger$, this unitarity request would suggest, of course, to use  $P_{\Phi_0\rightarrow\Phi_f}^\varphi$ rather than $P_{\Phi_0\rightarrow\Phi_f}^\Psi$. But, again, this is not really our main criterion.

\vspace{2mm}

We begin our analysis with a simple situation: let us assume  that $\Sc$ is prepared in the following linear combination: $\Phi_0=\varphi_a+\varphi_b$, where $0\leq a\leq N$, $0\leq b\leq N$, and $a\neq b$. Then $\Phi(t)=e^{-iHt}\Phi_0=e^{-iE_at}\varphi_a+e^{-iE_bt}\varphi_b$. Hence we get
$$
\|\Phi(t)\|^2=\|\varphi_a\|^2+\|\varphi_b\|^2+\left(e^{i(E_a-E_b)t}\left<\varphi_a,\varphi_b\right>+c.c.\right),
$$
$$
\|\Phi(t)\|_\Psi^2=2,
$$ and
$$
\|\Phi(t)\|_\varphi^2=\|\varphi_a\|_\varphi^2+\|\varphi_b\|_\varphi^2+\left(e^{i(E_a-E_b)t}\left<\varphi_a,\varphi_b\right>_\varphi+c.c.\right),
$$
where $c.c.$ stands for complex conjugate.

If we now take $\Phi_f=\varphi_j$, for some $j=0,1,2,\ldots,N$, we conclude first that
\be
P_{\Phi_0\rightarrow\Phi_f}^\Psi(t)=\frac{1}{2}\left(\delta_{j,a}+\delta_{j,b}\right).
\label{35}\en
Hence, according to this rule, a transition between $\Phi_0=\varphi_a+\varphi_b$ and $\varphi_j$ is possible if and only if $j=a$ or $j=b$. In all other cases,  a similar transition would not be allowed. However, this is not the same conclusion we get considering the other possible definitions. In fact we find
\be
P_{\Phi_0\rightarrow\Phi_f}(t)=\frac{|\left<\varphi_j,\varphi_a\right>|^2+|\left<\varphi_j,\varphi_b\right>|^2+\left(e^{i(E_a-E_b)t}\left<\varphi_a,\varphi_j\right>\left<\varphi_j,\varphi_b\right>+c.c\right)}{
\|\varphi_j\|^2\left(\|\varphi_a\|^2+\|\varphi_b\|^2+\left(e^{i(E_a-E_b)t}\left<\varphi_a,\varphi_b\right>+c.c\right)\right)}
\label{36}\en
and
\be
P_{\Phi_0\rightarrow\Phi_f}^\varphi(t)=\frac{|\left<\varphi_j,\varphi_a\right>_\varphi|^2+|\left<\varphi_j,\varphi_b\right>_\varphi|^2+\left(e^{i(E_a-E_b)t}\left<\varphi_a,\varphi_j\right>_\varphi
\left<\varphi_j,\varphi_b\right>_\varphi+c.c\right)}{
\|\varphi_j\|_\varphi^2\left(\|\varphi_a\|_\varphi^2+\|\varphi_b\|_\varphi^2+\left(e^{i(E_a-E_b)t}\left<\varphi_a,\varphi_b\right>_\varphi+c.c\right)\right)}.
\label{37}\en
It is clear that, if we perform an experiment on $\Sc$, and we find that the transition probability from a vector $\Phi_0=\varphi_a+\varphi_b$ to $\varphi_j$, with $j\neq a,b$, is different from zero, $P_{\Phi_0\rightarrow\Phi_f}^\Psi(t)$ must be discharged, while the other two possible definitions could still be correct, in principle.

In a similar way, if we fix now $\Phi_f=\Psi_j$ leaving unchanged $\Phi_0$, we deduce that
\be
P_{\Phi_0\rightarrow\Phi_f}(t)=\frac{\delta_{j,a}+\delta_{j,b}}{
\|\Psi_j\|^2\left(\|\varphi_a\|^2+\|\varphi_b\|^2+\left(e^{i(E_a-E_b)t}\left<\varphi_a,\varphi_b\right>+c.c\right)\right)},
\label{38}\en
\be
P_{\Phi_0\rightarrow\Phi_f}^\Psi(t)=\frac{|\left<\Psi_j,\Psi_a\right>|^2+|\left<\Psi_j,\Psi_b\right>|^2+\left(e^{i(E_a-E_b)t}\left<\Psi_a,\Psi_j\right>\left<\Psi_j,\Psi_b\right>+c.c\right)}{
2\|\Psi_j\|_\Psi^2},
\label{39}\en
and
\be
P_{\Phi_0\rightarrow\Phi_f}^\varphi(t)=\frac{|\left<\varphi_j,\varphi_a\right>|^2+|\left<\varphi_j,\varphi_b\right>|^2+\left(e^{i(E_a-E_b)t}\left<\varphi_a,\varphi_j\right>\left<\varphi_j,\varphi_b\right>+c.c\right)}{
\|\varphi_a\|_\varphi^2+\|\varphi_b\|_\varphi^2+\left(e^{i(E_a-E_b)t}\left<\varphi_a,\varphi_b\right>_\varphi+c.c\right)}.
\label{310}\en
We see that, with these choices of initial and final states, we should get a zero transition probability if $j\neq a,b$, at least if the correct transition law is given by
$P_{\Phi_0\rightarrow\Phi_f}(t)$. Suppose then that, making an explicit experiment and choosing $j\neq a,b$, we observe a  zero transition probability from $\Phi_0$ to $\Phi_f$. Then we are forced to assume that the only possible correct expression for such a probability is exactly the one given by $P_{\Phi_0\rightarrow\Phi_f}(t)$. Summarizing,  the probability functions proposed here all make perfect sense, and only some experiment can discriminate between them.

 We will say more in the next section, where an example is discussed in many details.

\vspace{2mm}

{\bf Remarks:--} (1) It remains open the case in which some of the eigenvalues of $H$ are complex. In this case we have already seen that problems may arise in the general settings proposed here, so a deeper analysis is required. Notice also that none of the formulas above describe damping: only oscillations are allowed! The reason is clear: our analysis, here, is restricted to real eigenvalues. In other words, even if the hamiltonian is non self-adjoint, its eigenvalues are real numbers anyhow. In order to get some decay, we need also some eigenvalues with non zero imaginary parts.

(2) In principle, we could repeat the same analysis using $H^\dagger$ instead of $H$, but we will not do it here. In fact, we do not expect any essential difference. On the contrary, we expect that the conclusion would be quite similar: we can still consider three possibilities as in (\ref{34}), replacing $H$ with $H^\dagger$,  but only experimental results could say which one is correct.

(3) Once we have been able to decide which one is the probability transition which is in agreement with the experiments, it is clear that we have a {\em preferred scalar product}, and our suggestion is that this is the one to be used in the computation, say, of the norm of the wave-function, in the determination of the adjoint of the operators (and of the observables in particular), and so on.

\section{A detailed example}

The system we will discuss here was considered recently in \cite{santos}, and it is described, in its simplified version, by the non self-adjoint hamiltonian
$$
H_{SDS}=- g \left(
                   \begin{array}{cc}
                     0 & 1-k \\
                     1+k & 0 \\
                   \end{array}
                 \right),
$$
where $g\in\Bbb R$ and $k\in]-1,1[$, and we have put $\hbar=1$ to simplify the notation. Of course, the interesting situation is when $k\neq0$, since otherwise $H_{SDS}=H_{SDS}^\dagger$. This can be written in terms of pseudo-fermionic operators, see Appendix, by introducing
$$
a=\frac{1}{2}\left(
                   \begin{array}{cc}
                     1 & 1/\alpha \\
                     -\alpha & -1 \\
                   \end{array}
                 \right),\qquad b=\frac{1}{2}\left(
                   \begin{array}{cc}
                     1 & -1/\alpha \\
                     \alpha & -1 \\
                   \end{array}
                 \right),
$$
where $\alpha=\sqrt{\frac{1+k}{1-k}}$. Then $N=ba=\frac{1}{2}\left(
                   \begin{array}{cc}
                     1 & 1/\alpha \\
                     \alpha & 1 \\
                   \end{array}
                 \right)$, and, taking $\rho=- g\sqrt{1-k^2}$ and $\omega=2\rho$, we deduce that $H_{SDS}=\omega N+\rho\1$. The eigenvectors of $H_{SDS}$, and of its adjoint, can now easily deduced:
$$
\varphi_0=N_\varphi\left(
                     \begin{array}{c}
                       1 \\
                       -\alpha \\
                     \end{array}
                   \right), \, \varphi_1=b\varphi_0=N_\varphi\left(
                     \begin{array}{c}
                       1 \\
                       \alpha \\
                     \end{array}
                   \right), \quad \Psi_0=N_\Psi\left(
                     \begin{array}{c}
                       1 \\
                       -1/\alpha \\
                     \end{array}
                   \right), \, \Psi_1=a^\dagger\Psi_0=N_\Psi\left(
                     \begin{array}{c}
                       1 \\
                       1/\alpha \\
                     \end{array}
                   \right),
$$
where $N_\varphi\,\overline{N_\Psi}=\frac{1}{2}$, to guarantee that $\left<\varphi_k,\Psi_l\right>=\delta_{k,l}$, $k,l=0,1$. Then $H\varphi_k=E_k\varphi_k$, with $E_0=\rho=-E_1$, which are both real for the range of $k$ allowed. Analogously, we can explicitly check that  $H^\dagger\Psi_k=E_k\Psi_k$, $k=0,1$. A simple computation shows that $\sum_{k=0}^1|\varphi_k\left>\right<\Psi_k|=\sum_{k=0}^1|\Psi_k\left>\right<\varphi_k|=\1$, while
$$
S_\varphi=\sum_{k=0}^1|\varphi_k\left>\right<\varphi_k|=2|N_\varphi|^2\left(
                                                                        \begin{array}{cc}
                                                                          1 & 0 \\
                                                                          0 & \alpha^2 \\
                                                                        \end{array}
                                                                      \right),\quad S_\Psi=\sum_{k=0}^1|\Psi_k\left>\right<\Psi_k|=2|N_\Psi|^2\left(
                                                                        \begin{array}{cc}
                                                                          1 & 0 \\
                                                                          0 & 1/\alpha^2 \\
                                                                        \end{array}
                                                                      \right).
$$
Then, as expected, $S_\varphi=S_\Psi^{-1}=S_\varphi^\dagger$, and they are positive operators, with obvious positive square roots $S_\varphi^{1/2}$ and $S_\Psi^{1/2}$. Due to the fact that these are diagonal matrices, it is a particularly simple exercise to check that all the properties listed in the Appendix are indeed satisfied: $S_\varphi\Psi_k=\varphi_k$, $S_\Psi N=N^\dagger S_\Psi$, and so on. Also, we could introduce $c=S_\Psi^{1/2}aS_\varphi^{1/2}=\frac{1}{2}\left(
                   \begin{array}{cc}
                     1 & 1/\alpha \\
                     -1 & -1 \\
                   \end{array}
                 \right)$, $N_0=c^\dagger c=\frac{1}{2}\left(
                   \begin{array}{cc}
                     1 & 1/\alpha \\
                     1 & 1 \\
                   \end{array}
                 \right)$, and the o.n. vectors $e_0=S_\Psi^{1/2}\varphi_0=\sqrt{2}N_\varphi|N_\Psi|\left(
                                                                                                      \begin{array}{c}
                                                                                                        1 \\
                                                                                                        -1 \\
                                                                                                      \end{array}
                                                                                                    \right)
                 $, and $e_1=S_\Psi^{1/2}\varphi_1=\sqrt{2}N_\varphi|N_\Psi|\left(
                                                                                                      \begin{array}{c}
                                                                                                        1 \\
                                                                                                        1 \\
                                                                                                      \end{array}
                                                                                                    \right)
                 $, as well as a self adjoint hamiltonian $H_0$ similar to $H$ and to $H^\dagger$:
                 $$
                 H_0=S_\Psi^{1/2} HS_\varphi^{1/2}=\rho\left(\1-2N_0\right)=\rho\left(
                                                                                  \begin{array}{cc}
                                                                                    0 & -1 \\
                                                                                    -1 & 0 \\
                                                                                  \end{array}
                                                                                \right).
                 $$
The different scalar products we can introduce in $\Bbb C^2$, given $f=\left(
                                                                                \begin{array}{c}
                                                                                  f_0 \\
                                                                                  f_1 \\
                                                                                \end{array}
                                                                              \right)
$ and $g=\left(
                                                                                \begin{array}{c}
                                                                                  g_0 \\
                                                                                  g_1 \\
                                                                                \end{array}
                                                                              \right),
$
are $\left<f,g\right>=\overline{f_0}g_0+\overline{f_1}g_1$, $\left<f,g\right>_\Psi=2|N_\Psi|^2\left(\overline{f_0}g_0+\frac{1}{\alpha^2}\overline{f_1}g_1\right)$ and  $\left<f,g\right>_\varphi=2|N_\varphi|^2\left(\overline{f_0}g_0+\alpha^2\overline{f_1}g_1\right)$. Moreover, the {\em new} adjoints for $H$, other than $H^\dagger$, using (\ref{2add1}) are found to be
$$
H^\flat=S_\Psi H^\dagger S_\varphi=- g\left(
                                             \begin{array}{cc}
                                               0 & \frac{(1+k)^2}{1-k} \\
                                               \frac{(1-k)^2}{1+k} & 0 \\
                                             \end{array}
                                           \right),
$$
while
$$
H^\sharp=S_\varphi H^\dagger S_\Psi=- g\left(
                                             \begin{array}{cc}
                                               0 & 1-k \\
                                               1+k & 0 \\
                                             \end{array}
                                           \right).
$$
Now, it is clear that $H^\sharp=H$. Also, it is a simple exercise to show that $H^\dagger=(H^\dagger)^\flat$, as we have seen in Section 2 for general reasons.

\vspace{2mm}

{\bf Remark:--} the use of PFs here could be thought as not really essential. And in fact, it is not. However, as we have discussed in \cite{baggar}, it provides a sort of elegant and unifying language for many finite-dimensional systems previously introduced in the literature by several authors. All these systems share a somehow common structure: biorthogonal sets, lowering, raising and number-like operators, {\em nice} anti-commutation rules, and so on, and all these features are quite naturally described in terms of PFs.

\subsection{The dynamics and transition probabilities}

Let us now consider what happens if we assume that $H$ drives the dynamical behavior of the system $\Sc$ we are considering in this section, via the equation $i\dot\Phi(t)=H\Phi(t)$. In particular, following what we have discussed in Section \ref{sectTPandC}, we compute the different transition probabilities for different choices of $\Phi_0$ and $\Phi_f$. To avoid useless complications, from now on we fix $N_\varphi=N_\Psi=\frac{1}{\sqrt{2}}$.

To begin with, let us take $\Phi_0=\varphi_0$ and $\Phi_f=\Psi_1$. Then $\Phi(t)=e^{-i\rho t}\varphi_0$ and we get
$$
P_{\Phi_0\rightarrow\Phi_f}(t)=0, \qquad P_{\Phi_0\rightarrow\Phi_f}^\Psi(t)=P_{\Phi_0\rightarrow\Phi_f}^\varphi(t)=\frac{k^2}{k^2+1}.
$$
As we see, these are all constant in time. It is clear that, if in an experiment, we compute the transition probability from $\varphi_0$ to $\Psi_1$, we could conclude that $P_{\Phi_0\rightarrow\Phi_f}(t)$ is the correct definition of the probability only if the result of the experiment gives zero. Otherwise, if we get $\frac{k^2}{k^2+1}$, we are not in a position to choose between $P_{\Phi_0\rightarrow\Phi_f}^\Psi(t)$ and $P_{\Phi_0\rightarrow\Phi_f}^\varphi(t)$, since they coincide. Then we need a second experiment. In particular, we can repeat the same measure, assuming again that $\Phi_0=\varphi_0$, but asking what changes if we now take $\Phi_f=\varphi_1$. This is a better choice since we get three different results:
$$
P_{\Phi_0\rightarrow\Phi_f}(t)=k^2, \qquad P_{\Phi_0\rightarrow\Phi_f}^\Psi(t)=0,\qquad P_{\Phi_0\rightarrow\Phi_f}^\varphi(t)=\frac{4k^2}{(k^2+1)^2},
$$
so that, if $k\neq1$, a single measure would be enough to discriminate between the three definitions.

\vspace{2mm}

{\bf Remarks:--} (1) If we take $k=0$ all the functions above reduce to zero. This is expected since, in this case, $H$ becomes self-adjoint, and therefore $\varphi_k$ coincides  with $\Psi_k$, and $\F_\varphi$ becomes an orthonormal set. Hence, in both cases considered above, we are asking which is the possibility that a system, originally prepared in an eigenstate of $H$, evolves toward a different eigenstate of $H$. Of course, in absence of interactions this possibility is zero, and this is exactly our result.

(2) The explicit results above show that each transition probability assumes values in $[0,1]$, as it should.

\vspace{2mm}

Because of the special conditions considered here, we have deduced probabilities which are constant in time. In order to get time-depending probabilities, we need to consider different initial conditions on $\Sc$. Let us consider now $\Phi_0=\varphi_0+\varphi_1$. Hence $\Phi(t)=e^{-i\rho t}\varphi_0+e^{i\rho t}\varphi_1$. Let further the final state be $\Phi_f=\Psi_0$. Then, with simple computations, we get:
$$
P_{\Phi_0\rightarrow\Phi_f}(t)=\frac{1-k^2}{2\left(1-k\cos(2\rho t)\right)}, \quad P_{\Phi_0\rightarrow\Phi_f}^\Psi(t)=\frac{1+k^2+2k\cos(2\rho t)}{2\left(1+k^2\right)},
$$
and
$$
P_{\Phi_0\rightarrow\Phi_f}^\varphi(t)=\frac{1}{2}.
$$
The first obvious remark is that, when $k=0$, these three functions collapse to the same value, $\frac{1}{2}$. The reason is clear, see Remark (1) above. When $k\neq 0$, the probabilities are indeed different and again it should be possible to discriminate between them with some experiment.

Summarizing,  to deduce some internal coherence of the whole framework, we should be able to decide first which one, between the different transition probabilities, is the one in agreement with the experiments, since they all have essentially the same mathematical properties. This, we believe, is crucial for a deeper understanding of the theory.

\section{Conclusions}\label{sectconl}

After a general discussion on the dynamical problem generated by a non self-adjoint hamiltonian, we have seen some consequences of our choices in the computations of several, inequivalent, transition probabilities. With the help of a simple example, we have proposed a way to choose the correct settings to be used. This could be useful to shed some light on this kind of systems, and in fact this is the core of the paper: as we have already pointed out several times, many authors have discussed in recent years the {\em dynamical problem} associated to non self-adjoint hamiltonians. However, in our knowledge, not many attempts have been made to clarify which one, among all the possible definitions, is really compatible with experiments.

It is clear that the extension of our analysis to infinite dimensional Hilbert spaces is highly non trivial. In fact quite often the operators involved turn out to be unbounded. This has consequences on the domains of the various operators, on the nature of the biorthogonal sets of eigenvectors of $H$ and $H^\dagger$, on the inequivalent topologies which can be introduced in the Hilbert space of the system, and so on, \cite{bagbook}. However, the fact that $H$ has a purely point spectrum needs not to be true anymore, and this complicates quite a bit the treatment in concrete situations. Even from this point of view, the use of PFs appears to be a good choice, since it allows a natural extension to  infinite dimensional spaces in terms of pseudo-bosons, which share with PFs many properties, \cite{bagbook}.

We should also mention that the case in which some eigenvalues of $H$ have non zero imaginary parts is surely more complicated, especially in view of its extension to infinite dimensional Hilbert spaces. The reason has been discussed in Section \ref{sectnatear}. Hopefully, the analysis of this situation will be undertaken soon.

\section*{Acknowledgements}
The author would like to acknowledge  support from the
   Universit\`a di Palermo and from Gnfm.

 \appendix

 \section{\hspace{-.7cm}ppendix:  Something on PFs}

The starting point is a modification of the CAR $\{c,c^\dagger\}=c\,c^\dagger+c^\dagger\,c=\1$,
$\{c,c\}=\{c^\dagger,c^\dagger\}=0$, between two operators, $c$ and $c^\dagger$, acting on a two-dimensional Hilbert space $\Hil$. The CAR are
replaced here by the following rules: \be \{a,b\}=\1, \quad \{a,a\}=0,\quad \{b,b\}=0, \label{a1}\en where the interesting situation is when
$b\neq a^\dagger$. These rules automatically imply that a non zero vector, $\varphi_0$, exists in $\Hil$ such that $a\,\varphi_0=0$, and that a
second non zero vector, $\Psi_0$, also exists in $\Hil$ such that $b^\dagger\,\Psi_0=0$, \cite{bagpf}. In general $\varphi_0\neq\Psi_0$.

Let us now introduce the  non zero
vectors $ \varphi_1=b\varphi_0$ and $\Psi_1=a^\dagger \Psi_0$,  as well as the non self-adjoint operators $ N=ba$ and
$N^\dagger=a^\dagger b^\dagger$.  We also introduce the self-adjoint operators $S_\varphi$ and $S_\Psi$ via their action on a
generic $f\in\Hil$: $$ S_\varphi f=\sum_{n=0}^1\br\varphi_n,f\kt\,\varphi_n, \quad S_\Psi f=\sum_{n=0}^1\br\Psi_n,f\kt\,\Psi_n. $$
Hence we have:

\begin{enumerate}

\item $$ a\varphi_1=\varphi_0,\quad b^\dagger\Psi_1=\Psi_0.$$

\item $$ N\varphi_n=n\varphi_n,\quad \N\Psi_n=n\Psi_n,$$
for $n=0,1$.

\item If the normalization of $\varphi_0$ and $\Psi_0$ are chosen in such a way that $\left<\varphi_0,\Psi_0\right>=1$,
then $$ \left<\varphi_k,\Psi_n\right>=\delta_{k,n},$$ for $k,n=0,1$.

\item $S_\varphi$ and $S_\Psi$ are bounded, strictly positive, self-adjoint, and invertible. They satisfy
$$ \|S_\varphi\|\leq\|\varphi_0\|^2+\|\varphi_1\|^2, \quad \|S_\Psi\|\leq\|\Psi_0\|^2+\|\Psi_1\|^2,$$ $$ S_\varphi
\Psi_n=\varphi_n,\qquad S_\Psi \varphi_n=\Psi_n,$$ for $n=0,1$, as well as $S_\varphi=S_\Psi^{-1}$. Moreover, the following
intertwining relations $$ S_\Psi N=\N S_\Psi,\qquad S_\varphi \N=N S_\varphi,$$ are satisfied.

\end{enumerate}

The above formulas show that (i) $N$ and $\N$ behave essentially as fermionic number operators, having eigenvalues 0 and 1 ; (ii) their related eigenvectors
are respectively the vectors of $\F_\varphi=\{\varphi_0,\varphi_1\}$ and $\F_\Psi=\{\Psi_0,\Psi_1\}$; (iii) $a$ and $b^\dagger$ are lowering
operators for $\F_\varphi$ and $\F_\Psi$ respectively; (iv) $b$ and $a^\dagger$ are rising operators for $\F_\varphi$ and $\F_\Psi$
respectively; (v) the two sets $\F_\varphi$ and $\F_\Psi$ are biorthonormal; (vi) the operators $S_\varphi$ and
$S_\Psi$  are self-adjoint, bounded, invertible, with bounded inverse, and  map $\F_\varphi$ in $\F_\Psi$ and viceversa; (vii) $S_\varphi$ and $S_\Psi$ intertwine between operators which are not self-adjoint. Moreover, see
 \cite{bagpf,bagpf2},  $\F_\varphi$
and $\F_\Psi$ are automatically Riesz bases for $\Hil$.

\vspace{2mm}

{\bf Remark:--} For completeness we have to mention the paper by Bender and Klevansky, \cite{ben3}, where similar generalized anti-commutation rules were introduced, but with a different perspective.

\end{document}